# Femtosecond covariance spectroscopy

Jonathan Tollerud[a,b,1], Giorgia Sparapassi[a,b,1], Angela Montanaro[a,b], Shahaf Asban[c], Filippo Glerean[a,b], Francesca Giusti[a,b], Alexandre Marciniak[a,b], George Kourousias[b], Fulvio Billè[b], Federico Cilento[b], Shaul Mukamel[c], Daniele Fausti[a,b,d]

[a]Physics department, University of Trieste, Trieste, I-34127

[b]Elettra-Sincrotrone Trieste S.C.p.A., Trieste, I-34149

[c]Chemistry Department, University of California, Irvine, CA 92617

[d] Department of Chemistry, University of Princeton, Princeton, NJ 08544

[1]These authors contributed equally to this work.

Corresponding authors: Correspondence and requests for materials should be addressed to Shaul Mukamel, 433A Rowland Hall, University of California, Irvine, CA 92617, +19498247600, email: smukamel@uci.edu and to Daniele Fausti, via Alfonso Valerio 2, Trieste, I-34127, +390403758449, email: daniele.fausti@elettra.eu.

Abstract: The success of non-linear optics relies largely on pulse-to-pulse consistency. In contrast, covariance based techniques used in photoionization electron spectroscopy and mass spectrometry have shown that wealth of information can be extracted from noise that is lost when averaging multiple measurements. Here, we apply covariance based detection to nonlinear optical spectroscopy, and show that noise in a femtosecond laser is not necessarily a liability to be mitigated, but can act as a unique and powerful asset. As a proof of principle we apply this approach to the process of stimulated Raman scattering in α-quartz. Our results demonstrate how nonlinear processes in the sample can encode correlations between the spectral components of ultrashort pulses with uncorrelated stochastic fluctuations. This in turn provides richer information compared to the standard non-linear optics techniques that are based on averages over many repetitions with well-behaved laser pulses. These proof-of-principle results suggest that covariance based nonlinear spectroscopy will improve the applicability of fs non-linear spectroscopy in wavelength ranges where stable, transform limited pulses are not available such as, for example, x-ray free electron lasers which naturally have spectrally noisy pulses ideally suited for this approach.

Introduction: Noise, intrinsic to the measurement of any physical quantity, is normally seen as a limitation to eliminate. The desired signal-to-noise ratio is commonly reached by i) mitigating as much as possible the amount of experimental noise and ii) taking the mean of a large number (N) of repeated 'identical' measurements. From an alternative perspective, where every repetition is considered to be a measurement under different conditions, noise can become an asset and be exploited as a source of additional information[1,2,3,4]. In this case, since the measurements are performed under different conditions, the mean value loses significance and other statistical tools such as higher order moments are needed. If treated properly, noise can help clarify the interpretation of experiments[5] and even amplify signals as in stochastic resonance schemes[6].

Femtosecond nonlinear optical spectroscopy is ideally suited for an approach based on high order moments. In standard mean value nonlinear spectroscopies, the nonlinear signals are often extremely weak relative to the linear ones and complicated experimental layouts are required to separate them. Pulsed sources typically have several amplification stages, which naturally lead to significant noise in the output, further complicating the detection[7]. In order to deal with these challenges, significant effort and investment has gone into engineering stable laser sources, and the stability requirements have influenced experimental design and technique development[8-10]. Laser cost and experimental complexity has limited the adoption of many extremely useful but overly difficult techniques (such as multidimensional spectroscopy, fs-SRS, etc.).

Pioneering work by Lau and Kummrow[11,12] in the 1980s and 90s and more recently by the Turner et al.[13] have shown that temporally incoherent (up to ns) pulses can be used in place of transform limited fs pulses to perform various nonlinear spectroscopic studies[14,15] including CARS and two-dimensional electronic spectroscopy without compromising on the required fs time resolution. These approaches are different for the present study since they require that each set of pulses are identical copies, and use traditional multi-beam geometries and mean-value detection.

We can instead consider each laser pulse in a femtosecond nonlinear experiment as a measurement under new conditions rather than a repetition of the same experiment. The spectrum of a nonlinear signal in an N-wave mixing experiment depends on the product of the excitation fields[10,16], and will thus change with the pulse-to-pulse fluctuations in the laser. In the approach proposed here we show that in the presence of spectrally narrow pulse-to-pulse fluctuations, the nonlinear sample response imprints correlations between the spectral components within the laser bandwidth. After recording each unique optical signal, we use it to calculate the covariance between intensities at different frequencies, rather than averaging all the signals out as in mean value approaches. Evaluating the frequency difference between those spectral components whose covariance is different from zero, we retrieve the energy of the sample excited states that have interacted with the radiation and thus introduced the optical correlation. From this viewpoint, it is clear that the larger are the stochastic fluctuations in the optical pulse, the smaller are the correlations between the pulse components before the interaction, and the better the sample non-linear response is explored. Thus adding a spectrally uncorrelated (i.e. spectrally narrow) stochastic element to each excitation pulse is the key to improve the covariance based approach. There has recently been some effort in combining single-pulse detection with metrics beyond mean value for detection of signals at the shot-noise level[17,18,19], however so far there has been no demonstration of enhancement of uncorrelated noise at a classical level to improve detection of spectral correlations in a nonlinear experiment.

In this manuscript, we use covariance based measurements to study vibrational modes in a crystalline quartz sample via stimulated Raman scattering (SRS). When repeated many times with different unique noise realizations, the amplitude of the different spectral components within the pulse bandwidth that are separated by the sample phonon frequencies become correlated (see appendix B of the supplementary materials for a microscopic description of the experiment). We stress that all of these correlated fluctuations are averaged out or never resolved in mean-value measurements, whereas the frequency resolved covariance of the transmitted pulses contains valuable insight on the SRS processes, and the energies of the phonons involved. Importantly, the framework used here to reveal SRS could be generalized to other nonlinear optical techniques based both on table top and free electron laser sources[20]. X-ray FELs pose a particularly attractive possibility because they function based on self-amplified spontaneous emission which intrinsically leads to noisy pulses optimally suited to covariance based techniques.

Results: In the present demonstration of femtosecond covariance spectroscopy, ultrafast pulses from a regenerative amplifier (~ 40 fs) are transmitted through the sample and the spectrum of each transmitted pulse is detected using a spectrometer that consists of a grating and a fast photodiode array detector.

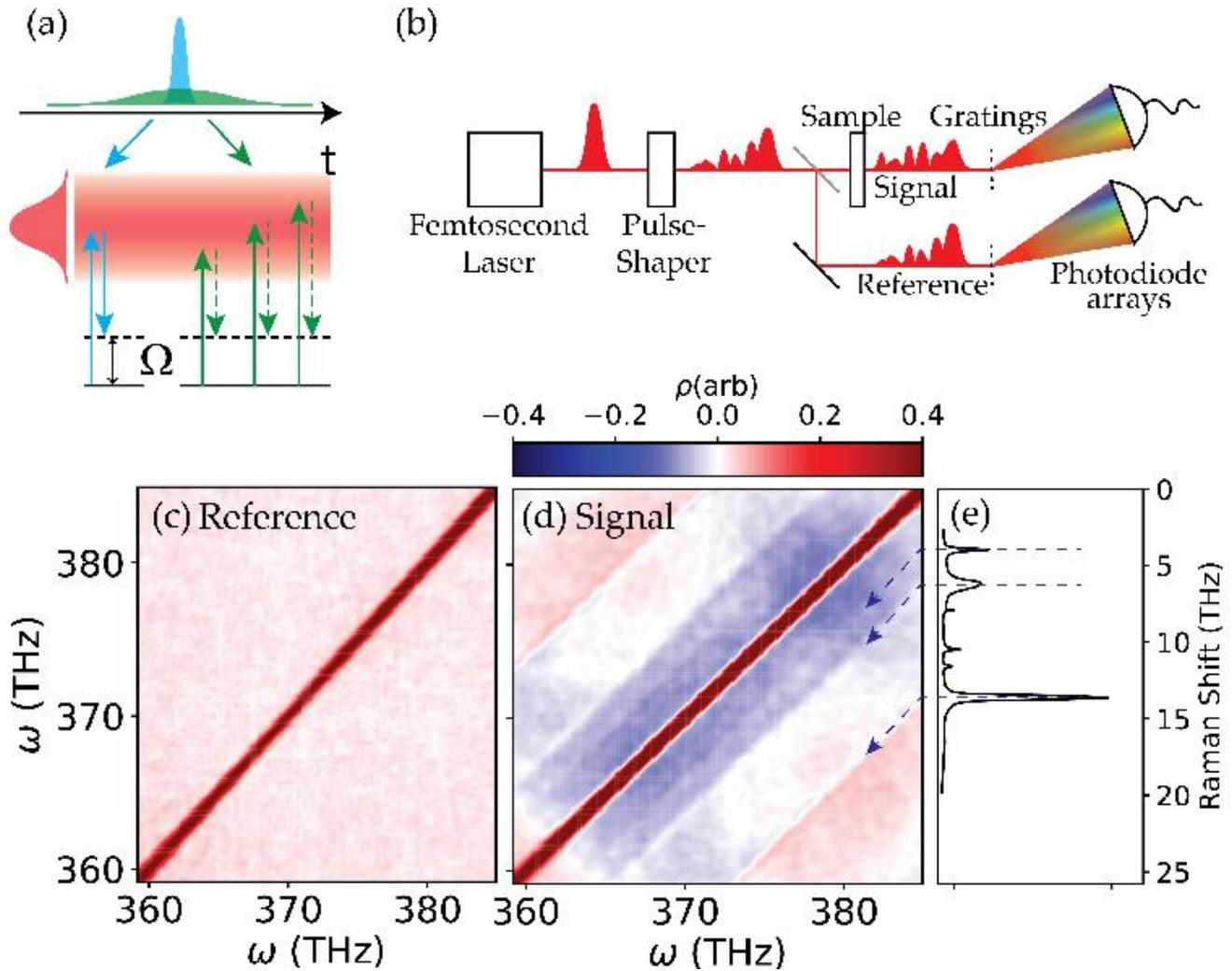

Figure 1 *(a) Simplified diagram of the ISRS process, which induces correlations of spectral components at frequencies $\omega$ and $\omega \pm \Omega$. The shaped pulses can be divided in a coherent component (blue) and a noisy tail (green). (b) The experimental apparatus. (c-d): Pearson correlation plots for all pairs of frequencies within the excitation spectrum (c) after the pulse shaper, and (d) after the quartz sample. (e) Spontaneous Raman spectrum of quartz[14], matching the positions of the features in the covariance map in (d).*

A stochastic element in the form of spectrally uncorrelated pulse-to-pulse noise can be added in several ways. We utilize a programmable liquid crystal spatial light modulator based pulse-shaper[21] placed between the laser and the sample as sketched in Fig. 1b. The pulse shaper changes the spectral phase with a defined modulation amplitude and correlation length ($\pm\pi/2$ RMS, 0.25 THz respectively). A reference beam is routed around the sample to a second identical spectrometer so we can compare the spectral covariance with and without interacting with the sample.

The time profiles resulting from the application of a spectral phase to initially transform limited pulses can be divided into a short coherent component (blue spike in Fig. 1a) which provides the impulsive excitation of the Raman modes, and noisy incoherent tails which probe the mode. Over the course of many noise realizations the average of the noisy tails becomes a roughly 1ps Gaussian pulse, shown in green in Fig. 1a.

In a typical measurement, we record sample and reference pulse spectra for 50,000 different noise realizations. We then use a covariance based analysis to extract information about the sample through correlations induced by SRS. While more diverse covariance metrics could be used, we consider the Pearson coefficient, which quantifies the degree of linear correlation between two random variables, the measured intensity I at the frequencies $\omega_i$ and $\omega_j$ within the pulse bandwidth:

$$P\left(I(\omega_i), I(\omega_j)\right) = \left[\langle I(\omega_i)I(\omega_j)\rangle - \langle I(\omega_i)\rangle\langle I(\omega_j)\rangle\right] / (\sigma_i\sigma_j)$$

where the angular bracket indicates a mean across all measurements, and $\sigma_{i(j)}$ is the standard deviation across all measurements of the intensity at frequency $\omega_{i(j)}$. P = 1(-1) indicates perfect correlation (anti-correlation), while P = 0 indicates no correlation. The result of this data processing performed across all the possible frequency combinations forms a 2D Pearson coefficient map, such as those shown in Fig 1c-d.

The P map calculated using the reference pulses, shown in Fig. 1c, exhibits no features apart from an area of positive correlation at $\omega_i = \omega_j$ (the diagonal of the map).
In contrast, when the pulse has interacted with the sample (Fig. 1d), the map is evidently structured. Most importantly, we observe signatures of correlation induced through SRS in the form of features offset of a quantity Δω from the diagonal, with a finite width which depends on the linewidth of the resonance and the correlation length. By comparing the correlation map to the spontaneous Raman spectrum[22] (Fig. 1e), it is clear that Δω matches the Raman shift of the main phonon features. The signal presence is substantiated by the fact that frequency components separated by Ω must have the same phase for interference between the paths leading to the population of that vibrational level to occur[23].

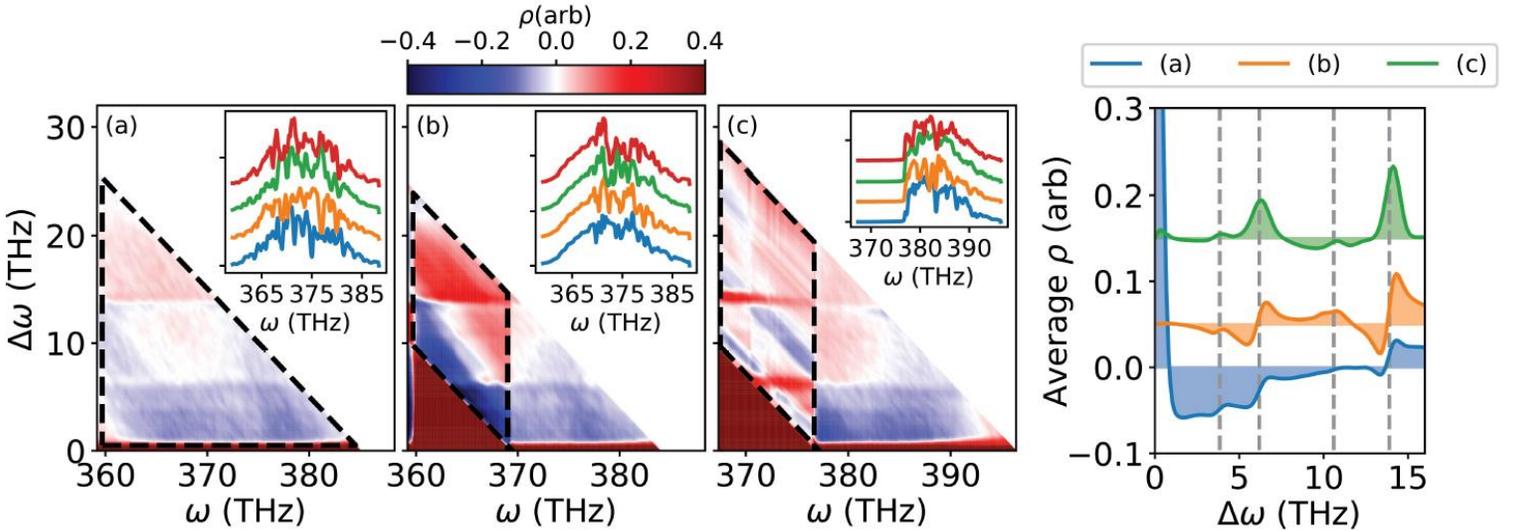

*Figure 2 Covariance maps recorded in α-quartz by applying uniform phase noise to each pulse. The noise is applied: (a) across the entire spectrum (b) only to frequencies above 369 THz (c) as in (b), but with mean value shaping to eliminate the frequencies below 369 THz from the pulse spectrum. The insets show some detected pulse spectra. (d) ISRS correlation spectra achieved by integrating along the ω axis the regions in (a)-(c) indicated by the black dashed lines.*

In order to visualize the Raman shift of the SRS features as a function of Δω, the correlation data can be plotted as P(ω,Δω) instead of P($\omega_i$, $\omega_j$)*. Fig. 2a shows this for the same data as Fig 1d. By taking the average Δω values of these maps along the ω axis, we can better analyze the features. In Fig. 2d, we show the average for the three regions indicated in Fig. 2a-c: this can be thought of as a sort of vibrational 'spectrum', where the Raman features appear on top of a shifting background.
The programmable nature of the pulse-shaper offers a broad flexibility in how the noise is introduced, which affects the visibility and the lineshape of the SRS features. Below, we explore two variations in which the noise is applied to only one half of the spectrum (above $\omega_0$ = 369 THz). In the first variation (shown in Fig. 2b), the spectral amplitude of the noise-free half of the spectrum (< $\omega_0$) is shaped such that the average value of the excitation spectrum has a Gaussian profile. In the second example (shown in Fig. 2c), the spectral amplitude of the noise-free side is reduced to 0, so that the average excitation spectrum has a sharp edge at $\omega_0$. In both cases, the features become more pronounced in the region quantifying the correlation between modulated and non-modulated components, indicated by the black dashed boxes in Fig. 2a-c. For comparison, the regions which quantify the correlation of two noisy frequencies have features similar to those achieved when the noise is applied uniformly across the entire spectrum.

---

* There is a symmetric set of features for negative Δω, which is left out of the plots in Fig. 2 for clarity purposes.

The resonances we identify appear at the spontaneous Raman resonances of quartz (which are indicated by the grey dashed lines)[22]. Clearly, the lineshape of these resonances depends on how the noise is applied. In most cases we observe a dispersive lineshape going from negative to positive values as $\Delta\omega$ increases, except in the case of the red region in Fig. 2c which has purely positive peak shapes.

Discussion: The Raman process shifts spectral weight, which leads to spectral correlations that manifest as off diagonal lines in the P-maps. The mechanism underlying this shift in spectral weight could have several origins, some coherent or incoherent. An SRS process produces a four-wave mixing signal which is coherent and – when spectrally and spatially overlapped with the excitation pulse – leads to self-heterodyning of the excitation pulse and SRS signal. The results in Fig 2 suggest that the process is coherent and stimulated; when there is a heterodyning field present at the emission energy (i.e. in Fig. 2 a, b) we observe dispersive peak shapes, but in the absence of a heterodyning field (Fig 2 c), we observe non-dispersive peak shapes. This dependence of the peak shape on the presence of a heterodyning field confirms the sensitivity of our detection technique to the signal phase, and that the mechanism of the spectral weight shift is coherent, consistent with SRS.

The dispersive peak shapes can be understood by considering the effect of self-heterodyning on the spectrally-resolved covariance. Regions where the signal and the excitation pulse are in phase should be positively correlated, as an increase in the excitation field will lead to a proportional increase in intensity at the signal frequency, due to constructive interference of the electric fields. Conversely when signals and the pulse are out of phase, their correlation is negative as the interference will be destructive. The SRS signal has a $\pi/2$ phase shift relative to the excitation pulse due to the absorption and reemission processes, which is consistent with the observed dispersive shapes.

It is important to note that the correlation maps are fundamentally different from the more familiar intensity spectra. For example, multiple mechanisms may induce correlations for overlapping pairs of frequencies. The shape of the resulting features in a P map is the sum of all these mechanisms, taking into account the sign of the correlations, which could be mistaken for interference of electric fields.

We also note that P is not necessarily directly proportional to the amplitude of the nonlinear signal: a weak signal can be strongly correlated to other frequencies. While not intuitive, this feature also illustrates the power of covariance-based detection: weak signals in mean-value measurements are easily masked by noise, but can be detected in the covariance plots, even when the spectra fully spectrally and spatially overlap with the excitation pulse.

In this work, we have shown that covariance based detection can be combined with noisy input pulses to resolve the SRS spectrum of a crystalline sample, and more incisive analytical tools than the average value can access a great depth of information that is missed by standard experiments. The transmitted pulses averaged over many noise realizations do not show the Raman resonances. However, the resonances are recovered in the covariance spectrum that reveals correlations between two spectral components separated by the phonon frequency. A high resolution Raman spectrum is thus generated by exploiting the noise fluctuations, without the need for stable light sources.

We emphasize that this result is only one example of a broader class of covariance-based analysis tools that can be applied both using table top sources, where fluctuations can be controlled, as well as using SASE Free Electron Lasers where amplitude and phase noise are unavoidable[24].

Further, while the two frequency Pearson coefficient based framework demonstrates how correlations induced by nonlinear processes can be detected, we expect that other tailored statistical tools will reveal non-linear processes at higher orders as done in multidimensional spectroscopy[25].

Acknowledgements: This work was supported by the European Commission through the ERC Starting Grant INCEPT (Grant. No 677488). S.M. gratefully acknowledges the support of National Science Foundation (Grant No CHE-1663822) and the Chemical Sciences, Geosciences, and Biosciences division, Office of Energy Sciences, Office of Science, U.S. Department of Energy through Award No. DE-FG02-04ER15571 and DE-SC0019484.. S.A. was supported by the DOE grant.

## Supplementary Information

**Experimental setup**
The experimental setup is shown in Fig. 1b of the main text. The source is an ultrafast laser delivering transform limited pulses of about 40 fs duration, with a Gaussian spectrum centered at 375 THz with a FWHM of ≈15 THz, and 2.4 mJ/PP. The pulses are produced by a Ti:Sapphire oscillator (Coherent Vitara) at 80 MHz and amplified using a 67 W pump laser at 5 kHz (Coherent Revolution) in an amplifier (Coherent Legend Duo) with two stages (regenerative cavity and single pass). The amplitude fluctuation of the laser output is about 5% shot to shot.

The pulse shaper uses a 512 x 512 pixel 2D liquid crystal on silicon Spatial Light Modulator (Meadowlark Optics) in reflection geometry[1]. It is set up for diffraction-based pulse shaping, which allows the phase-only liquid crystal matrix to simultaneously shape the spectral amplitude and phase[2]. The speed of the experiment is limited by the liquid crystals rotation time to be about 500 Hz. The beam is dispersed on the liquid crystal matrix so that the frequency band on a single pixel is about 0.1 THz. To control the resolution of the experiment, the random phase patterns we generate have a correlation length larger than the optical resolution of the pulse-shaper and the detector[3]. To achieve this, we first generate an array of 512 uniformly distributed random numbers. A Gaussian smoothing function with a standard deviation equal to the optical resolution is then used to smooth the array of random numbers. Finally, the resulting function is rescaled to achieve the desired RMS phase fluctuations (in this case $\pm\frac{\pi}{2}$). A reference pulse is created using a beam splitter between the pulse shaper and the sample. The sample beam is focused on a 1 mm thick crystalline $\alpha$-quartz sample, cut orthogonally to the microscopic c-axis plane, then collimated, the intensity can be controlled from 0 to 10 mJ/cm². The sample and reference beams are both detected using identical spectrometers consisting of a transmission diffraction grating, a 25 cm imaging lens (in 4-f geometry) and a 256 element linear silicon photodiode array (Hamamatsu).

**Model**
In this appendix we develop a model able to describe the role of the randomness of the phase in our covariance-based spectroscopic technique. We first consider the effect of such random fluctuations on the reference (non-interacting) beam; we then describe the Stimulated Raman Scattering (SRS) process and discuss how the stochasticity of the phase affects the measurement.

As discussed in the main text, each measurement of the frequency dispersed intensity is performed through a detector with finite frequency resolution. We model this by introducing a gate function $G(\omega - \bar{\omega})$, spanning a spectral region centered around the central frequency of the pixel $\bar{\omega}$. The spectral extension of the gating is set by the finite size of the pixel of the photodiode arrays. We can thus write the measured intensity on the pixel centered at $\bar{\omega}$ by considering the superposition of the fields within its spectral extension:

$$I(\bar{\omega}) = \Re \int d\omega' d\omega'' G(\omega' - \bar{\omega}) G(\omega'' - \bar{\omega}) E^*(\omega') E(\omega'') e^{-i[\varphi(\omega') - \varphi(\omega'')]} \qquad (1)$$

where $\varphi(\omega^{',''})$ is the stochastic phase of the incident pulses and $\omega^{',''}$ are the integration variables to cover the pixel size. In the experiments reported in the manuscript the stochastic phase introduced between different components is decaying with a characteristic frequency-scale $\Delta_{corr}$ (parameter controlled experimentally). The symbol $\langle\ldots\rangle$ represents the average over repeated measurements. We assume that the field amplitude $E(\omega)$ and the phase $\varphi(\omega)$ change slowly with $\varphi(\omega')$, i.e. $\Delta_{corr}$ is larger than the detector pixel size. We thus consider only the value of the fields at $\bar{\omega}$ and expand the phase to first order around $\bar{\omega}$ as

$$\varphi(\omega'') - \varphi(\omega') \approx (\omega'' - \omega') \frac{\partial \varphi}{\partial \omega}\Big|_{\bar{\omega}} = \Phi(\bar{\omega}) \qquad (2)$$

It should be noted that this treatment is not limited to Gaussian correlation functions only. It can be applied to any kind of decaying correlation functions whose scale is defined by a characteristic length.

By considering the expansion in Eq. (2), the overall intensity measured on the pixel is given by:

$$I(\bar{\omega}) = \Re\, E^*(\bar{\omega}) E(\bar{\omega}) e^{i\Phi(\bar{\omega})} = |E(\bar{\omega})|^2 \cos(\Phi(\bar{\omega})) \equiv \bar{I}(\bar{\omega}) \cos(\Phi(\bar{\omega})) \qquad (3)$$

It is worth to note that in this limit the phase changes slowly over the pixel size and therefore $\Phi(\overline{\omega}) \ll 1$.
We can now write the correlator between the intensities measured at different pixels with central frequencies $\overline{\omega}_{1,2}$ as:

$$\langle I(\overline{\omega}_1)I(\overline{\omega}_2)\rangle = \langle \bar{I}(\overline{\omega}_1)\cos(\Phi(\overline{\omega}_1))\bar{I}(\overline{\omega}_2)\cos(\Phi(\overline{\omega}_2))\rangle \tag{4}$$

The correlation function between the different spectral phases $\varphi(\omega_{1,2})$ results in a finite correlation length between their derivatives $\Phi(\omega_1)$ and $\Phi(\omega_2)$, resulting in an intensity correlator:

$$\langle I(\overline{\omega}_1)I(\overline{\omega}_2)\rangle = \begin{cases} \bar{I}(\overline{\omega}_1)\bar{I}(\overline{\omega}_2)\kappa_{12}^2, & if\ |\overline{\omega}_1 - \overline{\omega}_2| \lesssim \Delta_{corr} \\ 0, & if\ |\overline{\omega}_1 - \overline{\omega}_2| \gg \Delta_{corr} \end{cases} \tag{5}$$

where $\kappa_{12}^2 \equiv \langle \cos(\Phi(\overline{\omega}_1))\cos(\Phi(\overline{\omega}_2))\rangle$. Note that for frequency components farther apart than the correlation length, i.e. for $|\overline{\omega}_1 - \overline{\omega}_2| \gg \Delta_{corr}$ the product $\langle\cos(\Phi(\overline{\omega}_1))\cos(\Phi(\overline{\omega}_2))\rangle$ can be factorized in $\langle\cos(\Phi(\overline{\omega}_1))\rangle\langle\cos(\Phi(\overline{\omega}_2))\rangle = 0$ since $\Phi(\overline{\omega}_{1,2})$ are independent random variables with null average.
This mechanism maps phase fluctuations into amplitude ones and results in the Pearson coefficient map of the reference channel shown in Fig. 1c in the manuscript. As discussed in the main text and detailed here, $\Delta_{corr}$ sets the spectral resolution of the method proposed.

We stress that the crucial step in this approach is to consider the finite spectral resolution of the detectors[*]. Indeed, in the ideal case of a monochromatic detection, the random phase fluctuations would have no effects on the measured intensities, and no detectable correlations neither phase variations (Eq. (2)) would be expected within the reference beam.

In the following, we adopt a fully quantum field model[4] to calculate the optical signal. We describe the Stimulated Raman Scattering (SRS) process through a diagrammatic representation[4] and recast the signal in terms of transition amplitudes.
The detected signal of a quantum field is computed as the net time variation of the number of photons in the self-heterodyned transmitted field:

$$S \equiv \int dt \left[\frac{d}{dt}(\widehat{\mathcal{N}})_\rho\right] = \frac{i}{\hbar}\int [\widehat{\mathcal{H}}_{int}, \widehat{\mathcal{N}}] \tag{6}$$

where $\widehat{\mathcal{N}} \equiv \sum_s \hat{a}_s^\dagger \hat{a}_s$ and the symbol $(...)_\rho$ denotes the average over the density matrix operator of the whole system (field and matter). We have denoted by $\widehat{\mathcal{H}}_{int}$ the interaction Hamiltonian within the Rotating Wave Approximation, namely

$$\widehat{\mathcal{H}}_{int} = \hat{\mathcal{E}}(t)\hat{V}^\dagger + \hat{\mathcal{E}}^\dagger(t)\hat{V} \tag{7}$$

The commutator in Eq. (6) can be calculated by evaluating the canonical commutation relations for the bosonic operators. The solution of the Liouville-Von Neumann equation gives the time evolution of the density matrix in the interaction picture, so that the average can be now performed over the density matrix of the non-interacting system, denoted by $\langle ...\rangle$[5]. The signal can be recast in the following expression:

$$S = \frac{2}{\hbar}\mathfrak{I}\int dt\ \langle \hat{\mathcal{E}}_L^\dagger(t)\hat{V}_L(t)\ \mathcal{T}\ e^{-\frac{i}{\hbar}\int_{-\infty}^t d\tau\ H_{int-}(\tau)}\rangle \tag{8}$$

where $\mathcal{T}$ is the time-ordering operator in the Liouville space and $H_{int-}(\tau)$ is the time-dependent commutation relation with the interaction Hamiltonian in Eq. (7). We have adopted the L/R representation of the Liouville superoperators introduced in[4]. The electric field operator and the electric dipole one, are respectively defined as:

$$\hat{E}(t) = \hat{\mathcal{E}}(t) + \hat{\mathcal{E}}^\dagger(t) \tag{9}$$

$$\hat{\mu}(t) = \hat{V}(t) + \hat{V}^\dagger(t) \tag{10}$$

---

[*] Note that, even if we have discussed here only the role of the detectors, the results we have obtained can be extended to all non-ideal coarse-grained instruments which are responsible for summations over neighboring spectral modes. It is likely that also the SLM introduces similar effects on the pulses.

Since we detect intense fields, i.e. fields in a classical regime, we replace in the following the electric field operators $\hat{\mathcal{E}}(t)$ and $\hat{\mathcal{E}}^\dagger(t)$ with their expected values $\mathcal{E}(t)$ and $\mathcal{E}^*(t)$. Moreover, when the coupling between the field modes and the matter is off resonances, the equality $V(t) = \alpha(t)\mathcal{E}(t)$ holds. The signal can be then expressed in terms of the polarizability as:

$$S = \frac{2}{\hbar} \Im \int dt \, \langle \mathcal{E}_L^*(t)\mathcal{E}_L(t)\alpha_L(t) \, \mathcal{T} \, e^{-\frac{i}{\hbar}\int_{-\infty}^{t} d\tau \, H_{int-}(\tau)} \rangle \tag{11}$$

Since we perform a frequency-resolved shot-to-shot detection of the pulses transmitted by the sample, we are interested in the frequency dispersed signal $S(\omega)$. We can then consider a frequency gating $\delta(\omega - \bar{\omega})$ and, by Fourier transform the electric field, get:

$$S(\omega) = \frac{2}{\hbar} \Im \int dt \, e^{i\omega t} \langle \mathcal{E}_L^*(\omega)\mathcal{E}_L(t)\alpha_L(t) \, \mathcal{T} \, e^{-\frac{i}{\hbar}\int_{-\infty}^{t} d\tau \, H_{int-}(\tau)} \rangle \tag{12}$$

Note that the $0^{th}$-order in the last equation contains one light-matter interaction which results in a vanishing trace. We thus expand to the first (nontrivial) order as:

$$S(\omega) = \frac{2}{\hbar} \Im \int dt \, e^{i\omega t} \langle \mathcal{E}_L^*(\omega)\mathcal{E}_L(t)\alpha_L(t) \left(-\frac{i}{\hbar}\right)\int_{-\infty}^{t} d\tau \, H_{int-}(\tau) \rangle \tag{13}$$

The signal thus splits into two terms: the first one ($S_a$) involves interactions from both the left and the right, the second one ($S_b$) only interactions from the left.

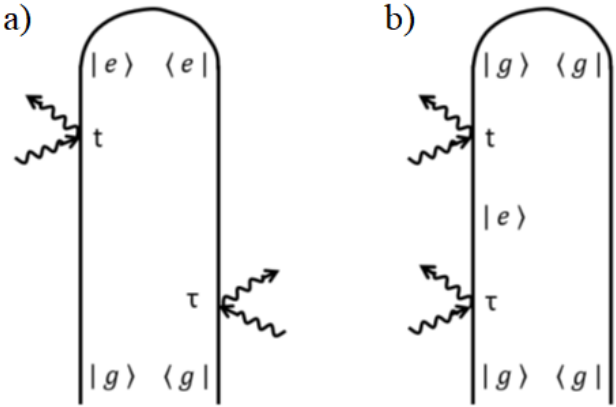

*Fig. S1.* *Closed-time-path-loop diagrams describing the Stimulated Raman Scattering process. a) Stokes process. b) Anti-Stokes process.*

We can diagrammatically describe them by using a Closed-Time-Path-Loop (CTPL) representation, following the rules given in[4]. The CTPL diagrams (Fig. S1) represent the process described by the latter equation. The diagram on the left represents the energy loss of the electric field (Stokes process), while the one on the right represents the energy gain of the electric field (Anti-Stokes process).
First we explicitly evaluate $S_a$:

$$S_a(\omega) = \frac{2}{\hbar^2} \Im \int\int \frac{d\omega_1 d\omega_2 d\omega_3 d\omega_4 d\omega_5}{(2\pi)^5} \frac{\langle \mathcal{E}^*(\omega)\mathcal{E}(\omega_1)\mathcal{E}^*(\omega_3)\mathcal{E}(\omega_4) \rangle}{-\omega_3 + \omega_4 + \omega_5 - i\gamma} \langle \alpha_L(\omega_2)\alpha_R(\omega_5) \rangle_\mu \, 2\pi \, \delta(\omega_1 + \omega_2 + \omega_4 + \omega_5 - \omega_3 - \omega) \tag{14}$$

where in the last step we have explicitly written the interaction Hamiltonian and Fourier transformed both the fields and the polarizabilities. We have then explicitly written the material degrees of freedom, so that now the polarizabilities can be spanned in the energy eigenstates of the material.

We consider here a two-level system (the ground state is denoted by $|g\rangle$, while the vibrational excited state by $|e\rangle$) with a single vibrational frequency $\Omega$. The term involving the polarizability is then given by:

$$\langle \alpha_L(\omega_2)\alpha_R(\omega_5)\rangle_\mu = \langle e|\alpha(\omega_2)|g\rangle \langle g|\alpha(\omega_5)|e\rangle = |\alpha_{ge}|^2 (2\pi)^2 \delta(\omega_2+\Omega)\delta(\omega_5-\Omega) \tag{15}$$

where in the last step we have expressed the average in the Kramers-Heisenberg form (a generalized Fermi golden rule), as in[6]. We have denoted by $|\alpha_{ge}|^2$ the polarizability transition amplitude between the two states.

By solving the Dirac deltas, we get the following expression for $S_a$:

$$S_a(\omega) = \frac{2|\alpha_{ge}|^2}{\hbar^2} \Im \iint \frac{d\omega_1 d\omega_2}{(2\pi)^2} \frac{\langle \mathcal{E}^*(\omega)\mathcal{E}(\omega+\omega_1-\omega_2)\mathcal{E}^*(\omega_1)\mathcal{E}(\omega_2)\rangle}{-\omega_1+\omega_2+\Omega-i\gamma} \tag{16}$$

The term $S_b$ can be evaluated in a similar fashion, with the exception that now the polarizability reads:

$$\langle \alpha_L(\omega_2)\alpha_L(\omega_5)\rangle_\mu = \langle g|\alpha(\omega_2)|e\rangle \langle e|\alpha(\omega_5)|g\rangle = |\alpha_{ge}|^2 (2\pi)^2 \delta(\omega_2-\Omega)\delta(\omega_5+\Omega) \tag{17}$$

By recombining the two terms, we get:

$$S(\omega) = \frac{2|\alpha_{ge}|^2}{\hbar^2} \Im \iint \frac{d\omega_1 d\omega_2}{(2\pi)^2} \mathcal{E}^*(\omega)\mathcal{E}(\omega+\omega_1-\omega_2)\mathcal{E}^*(\omega_1)\mathcal{E}(\omega_2) \left[\frac{1}{-\omega_1+\omega_2+\Omega-i\gamma} - \frac{1}{-\omega_1+\omega_2-\Omega-i\gamma}\right] \tag{18}$$

We can now use the Sokhotski-Plemelj theorem to explicitly calculate the terms within the square brackets. The theorem states that:

$$\frac{1}{\omega-i\gamma} = \mathcal{PP}\frac{1}{\omega} + i\pi\delta(\omega) \tag{19}$$

Since $\mathcal{E}^*(\omega)\mathcal{E}(\omega+\omega_1-\omega_2)\mathcal{E}^*(\omega_1)\mathcal{E}(\omega_2)$ is smooth and even around the poles of the integrand function, the principal values vanish when the integration over $\omega_{1,2}$ is performed. Therefore, by considering the action of the Dirac deltas, we get:

$$S(\omega) = \frac{|\alpha_{ge}|^2}{\hbar^2} \Re \iint \frac{d\omega'}{2\pi} [\mathcal{E}^*(\omega)\mathcal{E}(\omega+\Omega)\mathcal{E}^*(\omega')\mathcal{E}(\omega'-\Omega) - \mathcal{E}^*(\omega)\mathcal{E}(\omega-\Omega)\mathcal{E}^*(\omega')\mathcal{E}(\omega'+\Omega)] \tag{20}$$

So far, we have neglected the temperature dependence of the system. If the sample has a finite temperature, two additional processes should be considered, since the system can be initially either in the ground state or in the vibrationally excited one. The additional two contributions we get are identical to $S_a$ and $S_b$ if one replaces $\Omega \to -\Omega$. We thus get the same contributions with a minus sign. If we assume the system to be at thermal equilibrium of inverse temperature $\beta$, the temperature dependence can be included by considering the thermal distributions for the system, given by $p_g = \frac{1}{1+e^{-\beta\hbar\Omega}}$ for the ground state and by $p_e = \frac{1}{1+e^{\beta\hbar\Omega}}$ for the excited one. The final signal reads:

$$S(\omega;\Gamma) = \frac{|\alpha_{ge}|^2 p_{ge}(\beta)}{\hbar^2} \Re \int \frac{d\omega'}{2\pi} [\mathcal{E}^*(\omega)\mathcal{E}(\omega+\Omega)\mathcal{E}^*(\omega')\mathcal{E}(\omega'-\Omega) - \mathcal{E}^*(\omega)\mathcal{E}(\omega-\Omega)\mathcal{E}^*(\omega')\mathcal{E}(\omega'+\Omega)] \tag{21}$$

where we have defined the factor $p_{ge}(\beta) = p_g - p_e$ and introduced the parameter $\Gamma$, which includes all the field parameters that can be tuned in the experiment.

We consider a frequency dependent stochastic phase $\varphi(\omega)$ with frequency scale $\Delta_{corr}$, such that the phase correlations are considered statistically orthogonal $\langle \varphi(\omega_1)\varphi(\omega_2)\rangle = 0$ when they are far enough and $|\omega_1-\omega_2| \gg \Delta_{corr}$. Since the phases of two spectral components are correlated only if their frequency difference is smaller than $\Delta_{corr}$, this quantity sets the phase stochasticity scale. This additional frequency scale plays a key role in setting the spectral resolution while considering the signal correlation function. Assuming $\Omega \gg \Delta_{corr}$, one may regard $\Phi(\omega)$ and $\Phi(\omega+\Delta_{corr})$ as statistically independent variables.

Under these conditions, we can rewrite the signal making explicit the spectral phases $\varphi(\omega)$. By grouping the phase factors of each four-field product and taking the real part of the above equation, we get:

$$S(\omega; \Gamma) = \frac{|\alpha_{ge}|^2 p_{ge}(\beta)}{\hbar^2} \int \frac{d\omega'}{2\pi} [E^*(\omega)E(\omega + \Omega)E^*(\omega')E(\omega' - \Omega)\cos(\gamma) - E^*(\omega)E(\omega - \Omega)E^*(\omega')E(\omega' + \Omega)\cos(\beta)] \quad (22)$$

where we have defined the following quantities:
$$\begin{cases} \gamma = \varphi(\omega + \Omega) - \varphi(\omega) + \varphi(\omega' - \Omega) - \varphi(\omega') \\ \beta = \varphi(\omega - \Omega) - \varphi(\omega) + \varphi(\omega' + \Omega) - \varphi(\omega') \end{cases}$$

Since we have assumed that $\Omega \gg \Delta_{corr}$, the average values of $\cos(\gamma)$ and $\cos(\beta)$ vanish (and so does the average value of the signal) unless $\cos(\gamma)$ and $\beta$ are both zero, i.e. unless

$$\begin{cases} \omega' = \omega + \Omega & \text{for the first sum} \\ \omega' = \omega - \Omega & \text{for the second sum} \end{cases}$$

We can thus write the average transmitted signal as:

$$\langle S(\omega; \Gamma) \rangle = \frac{|\alpha_{ge}|^2 p_{ge}(\beta)}{2\pi\hbar^2} \{|E(\omega)E(\omega + \Omega)|^2 - |E(\omega)E(\omega - \Omega)|^2\} \quad (23)$$

which correctly describes the spectral (red or blue) shift due to the inelastic scattering[7]. We stress that when the pulse is very broad it is not possible to retrieve the Raman frequency $\Omega$ from the average signal.

To compute the cross-correlation signal $\langle S(\omega_i; \Gamma)S(\omega_j; \Gamma) \rangle$, we must evaluate the averages of the cosine products coming from the two integrals (whose integration variables are denoted with prime and double prime, respectively):

a)    $\langle \cos(\gamma_i') \cos(\gamma_j'') \rangle = \frac{1}{2} \langle \cos(\gamma_i' + \gamma_j'') + \cos(\gamma_i' - \gamma_j'') \rangle$

b)    $\langle \cos(\beta_i') \cos(\beta_j'') \rangle = \frac{1}{2} \langle \cos(\beta_i' + \beta_j'') + \cos(\beta_i' - \beta_j'') \rangle$

c)    $\langle \cos(\gamma_i') \cos(\beta_j'') \rangle = \frac{1}{2} \langle \cos(\gamma_i' + \beta_j'') + \cos(\gamma_i' - \beta_j'') \rangle$

d)    $\langle \cos(\beta_i') \cos(\gamma_j'') \rangle = \frac{1}{2} \langle \cos(\beta_i' + \gamma_j'') + \cos(\beta_i' - \gamma_j'') \rangle$

where we have denoted by the subscripts $i$ and $j$ the detected frequencies.
It is useful to consider the second term from equation c) and the first term from equation a):

i)    $\gamma_i' - \gamma_j'' = \varphi(\omega_i + \Omega) - \varphi(\omega_i) + \varphi(\omega' - \Omega) - \varphi(\omega') - [\varphi(\omega_j + \Omega) - \varphi(\omega_j) + +\varphi(\omega'' - \Omega) - \varphi(\omega'')]$

ii)    $\gamma_i' + \beta_j'' = \varphi(\omega_i + \Omega) - \varphi(\omega_i) + \varphi(\omega' - \Omega) - \varphi(\omega') + [\varphi(\omega_j - \Omega) - \varphi(\omega_j) + +\varphi(\omega'' + \Omega) - \varphi(\omega'')]$

These terms yield narrow distributions (delta-like) upon averaging. In particular, they give the following contractions:

i)    $\langle \cos(\gamma_i' - \gamma_j'') \rangle = \delta(\omega'' - \omega'') \delta(\omega_j - \omega_i)$

ii)    $\langle \cos(\gamma_i' + \beta_j'') \rangle = \delta(\omega' - \omega'' - \Omega) \delta(\omega_i - \omega_j + \Omega)$

Similar contractions arise also from the remaining terms.
The terms which contribute with the contractions of the type i) are trivial, since represent a contribution to the main diagonal of the Pearson coefficient map. Contractions of the type ii) are the interesting ones, since give rise to distinct off-diagonal sidebands shifted by the Raman frequency from the diagonal.

We explicitly calculate one of the terms of the second kind:

$$\begin{aligned}
\langle S(\omega_i,\Gamma)S(\omega_j,\Gamma)\rangle_{\gamma_i'+\beta_j''} &= \frac{|\alpha_{ge}|^4 p_{ge}^2(\beta)}{\hbar^4} \int \frac{d\omega'}{2\pi}\int \frac{d\omega''}{2\pi}\, \delta(\omega'-\omega''-\Omega)\,\delta(\omega_i-\omega_j+\Omega)\times \\
&\quad \times \left[E^*(\omega_i)E(\omega_i+\Omega)E^*(\omega')E(\omega'-\Omega)E^*(\omega_j)E(\omega_j-\Omega)E^*(\omega'')E(\omega''+\Omega)\right] = \\
&= \frac{|\alpha_{ge}|^4 p_{ge}^2(\beta)}{\hbar^4} |E(\omega_i)E(\omega_i+\Omega)|^2 \int \frac{d\omega'}{(2\pi)^2} |E(\omega')E(\omega'-\Omega)|^2\, \delta(\omega_i-\omega_j+\Omega)
\end{aligned} \quad (24)$$

The contribution above adds a distinct line which is shifted from the diagonal by the Raman frequency. Note that the condition of statistical dependence between two frequencies is satisfied for all the distances smaller or comparable to the distribution stochasticity scale, $\Delta_{corr}$. Hence, the width of the line (infinitesimal, in Eq. (24)) must be considered finite, leading to a blurring of the signal.